\documentclass[sigconf]{acmart}

\AtBeginDocument{%
  }

\copyrightyear{2026}
\acmYear{2026}
\setcopyright{cc}
\setcctype{by-nc-nd}
\acmConference[UbiComp Companion '26]{Companion of the 2026 ACM International Joint Conference on Pervasive and Ubiquitous Computing}{October 11--15, 2026}{Shanghai, China}
\acmBooktitle{Companion of the 2026 ACM International Joint Conference on Pervasive and Ubiquitous Computing (UbiComp Companion '26), October 11--15, 2026, Shanghai, China}
\acmDOI{10.1145/3798063.3837177}
\acmISBN{979-8-4007-2533-3/2026/10}

\begin{document}

\title{Mirroring the Past: Exploring How Ancestral Digital Self Influences History Learning}

\author{Duo Gong}
\authornote{Duo Gong and Fan Sun contributed equally to this work.}
\email{12333216@mail.sustech.edu.cn}
\affiliation{%
  \institution{Southern University of Science and Technology}
  \city{Shenzhen}
  \country{China}
}

\author{Fan Sun}
\authornotemark[1]
\email{12211327@mail.sustech.edu.cn}
\affiliation{%
  \institution{Southern University of Science and Technology}
  \city{Shenzhen}
  \country{China}
}

\author{Yucen Wang}
\email{12212852@mail.sustech.edu.cn}
\affiliation{%
  \institution{Southern University of Science and Technology}
  \city{Shenzhen}
  \country{China}
}

\author{Yufan Hu}
\email{huyufannju@163.com}
\affiliation{%
  \institution{Southern University of Science and Technology}
  \city{Shenzhen}
  \country{China}
}

\author{Wen Zhong}
\email{zhongw@mail.sustech.edu.cn}
\affiliation{%
  \institution{Southern University of Science and Technology}
  \city{Shenzhen}
  \country{China}
}

\author{Wei Zhang}
\email{124699938@qq.com}
\affiliation{%
  \institution{Shenzhen University}
  \city{Shenzhen}
  \country{China}
}

\author{Pengcheng An}
\authornote{Corresponding author.}
\email{anpc@sustech.edu.cn}
\affiliation{%
  \institution{Southern University of Science and Technology}
  \city{Shenzhen}
  \country{China}
}

\renewcommand{\shortauthors}{Gong et al.}

\renewcommand{\shortauthors}{Gong et al.}


\begin{abstract}

Learners often perceive history as distant from themselves, which limits immersion and empathy in history learning. To bridge this gap, we introduce the “Ancestral Digital Self”—an AI-generated pedagogical agent presented in prerecorded videos that mirrors the learner’s facial features and vocal timbre, representing a historically situated version of the self. We developed a reproducible workflow for creating AI-generated historical learning videos and conducted a within-subjects study ($N=36$) comparing a Digital Self agent with a non-self pedagogical agent. The Digital Self agent enhanced experiential measures, including narrative transportation, perceived relatedness, self–other inclusion, and agent perception. However, it did not improve immediate learning outcomes: quiz scores were lower in the Digital Self condition, and Remember/Know judgments showed no reliable differences. Interviews further suggested that self-similarity increased familiarity and motivation, while novelty and uncanniness could draw attention away from historical content. These findings offer design implications for future educational environments supported by pedagogical agents.

\end{abstract}

\begin{teaserfigure}
  \centering
  \includegraphics[width=\textwidth]{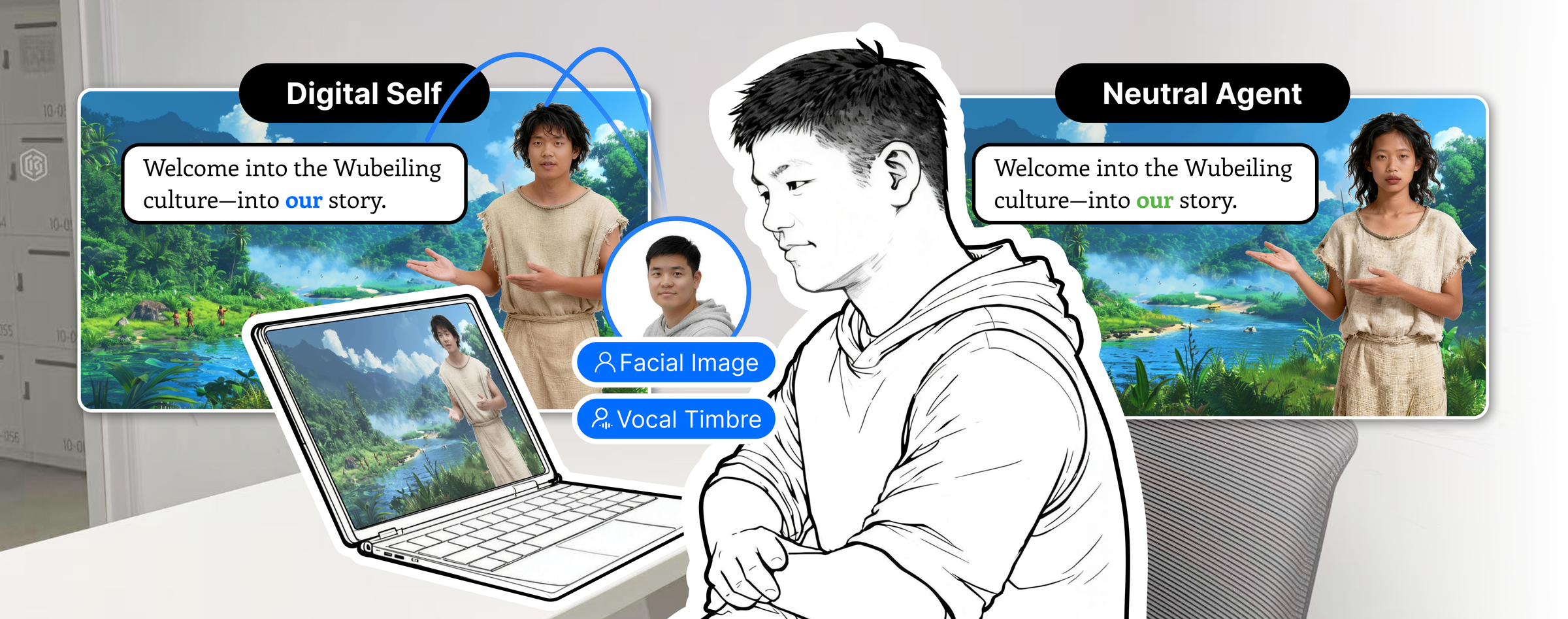} 
  \caption{The image illustrates a comparative experiment featuring two pedagogical agents narrating historical content related to the Wubeiling culture. Left: The digital self generated based on the user's own facial features and voice timbre. Right: A non-self pedagogical agent designed with reference to typical facial features in South China, where Wubeiling is located.}
  \Description{这里写图片的详细描述，用于辅助阅读}
  \label{fig:teaser}
\end{teaserfigure}




\maketitle

\section{Introduction}
History learning plays a significant role in facilitating both formal academic knowledge and informal civic literacy. Learners often struggle to understand historical actors’ thinking and actions, and historical narratives are frequently perceived as distant from their personal experiences, which can hinder immersion and emotional engagement  \cite{Thorp2020}. Prior work has shown that rote memorization is insufficient for historical learning \cite{mayer2002rote}. Presenting past events through vivid narratives helps learners understand history cognitively and emotionally \cite{endacott2013updated, 10.1145/1357054.1357291}.

In this study, we define the "Ancestral Digital Self" as an AI-generated pedagogical agent  in prerecorded videos that mirrors the learner’s facial features and vocal timbre, representing a historically situated version of the self. Unlike conventional pedagogical agents, it situates the learner’s identity within concrete historical contexts. Prior research indicates that high self-similarity can improve learning performance in various scenarios, such as training, public speaking, dance, and VR games \cite{ste2011feedforward,kao2022audio,clarke2023fakeforward,10.1145/3537972.3537991}.However, it is still unclear whether a digital self can enhance learning in historical contexts. This raises our research question: how does presenting the historical narrator as "another self" influence learners’ experiences and immediate learning outcomes?

To address this question, we embed the learner’s identity directly into historical scenarios. We designed a reproducible workflow in collaboration with an archaeology team: AI was used to generate historical materials, learners’ facial and vocal information was captured to create digital self narrators, and these narrators were integrated with the historical content using video editing software to produce standardized experimental videos. 

We conducted a within-subjects study with 36 participants, comparing the Ancestral Digital Self with a non-self pedagogical agent. Learning experiences were assessed across closeness (Inclusion of Other in the Self, IOS)\cite{aron1992inclusion}, agent perception (Agent Persona Instrument, API)\cite{baylor2003api}, relatedness (Intrinsic Motivation Inventory – Relatedness, IMI)\cite{mcauley1989psychometric}, narrative transportation (Narrative Transportation scale, NT)\cite{green2000role}, and uncanny responses (Uncanny Valley scale, UVS)\cite{ho2017measuring}. Immediate learning outcomes were evaluated through quizzes related to historical learning videos and the Remember/Know (R/K) paradigm\cite{gardiner1988functional}. The results showed that the digital self enhanced experiential measures while slightly increasing uncanny responses, but did not improve short-term learning outcomes. These findings provide design guidance for the use of digital selves in historical learning.

\section{Related Work}

Historical learning is often constrained by psychological distance and limited immersion, making it difficult for learners to situate themselves in historical contexts and understand historical actors’ intentions and actions \cite{Thorp2020}. Pedagogical agents (PAs) and AI-driven virtual characters have been used in narrative-centered learning to provide guidance and role-based simulations that support engagement, comprehension, and knowledge acquisition \cite{10.1145/3706598.3713109, ZHAO2025, baylor2003api}. However, because such agents are typically generic “strangers,” they may offer limited self-relevance and thus constrain identification, immersion, and affective connection with historical narratives \cite{MASSARA2013108, endacott2013updated}.

Video Self-Modeling (VSM) allows learners to observe themselves performing tasks, promoting engagement and skill refinement \cite{dowrick2012self}. VSM effects are theoretically grounded in the Self-Reference Effect  \cite{rogers1977self}, where self-related cues serve as salient anchors facilitating deeper cognitive and emotional processing. Recent work empirically demonstrates the Self-Reference Effect in generative AI contexts \cite{Kim2024}, updating the theoretical grounding beyond classic studies. Practical implementations include video replay, self-modeling videos, digital avatars, and VR environments \cite{Hiromitsu2024effects, Barathi2018interactive, Fitton2022dancing}, providing multimodal cues that enhance learner immersion and engagement.

History education agents have also been explored through interactive dialogue \cite{Park2025}. In contrast, our identity-mirroring approach embeds learners’ digital selves into historical narration videos, emphasizing experiential engagement through self-representation. Despite extensive research on VSM and digital self agents in skill learning, their application in historical learning remains underexplored. This gap motivates our study: embedding learners’ digital selves into historical narration videos to investigate impacts on learning experiences and outcomes.

\section{Methodology}


The historical learning content designed for this study centers on Wubeiling culture, a prehistoric culture dating back over 3,000 years in Shenzhen---the region where the study was conducted. We evaluated fixed, prerecorded AI-generated narration videos rather than a real-time interactive system; hereafter, \emph{agent} refers to the on-screen video-based pedagogical narrator.

\subsection{Participants}
We recruited 36 participants (21 males, 15 females; aged 19--29 years) from diverse educational backgrounds, including undergraduates (63.9\%), master's students (27.8\%), and others (8.3\%). Participants came from varied disciplines, including STEM (61.1\%), design and arts (25.0\%), and humanities and social sciences (13.9\%). Most participants reported no or little prior familiarity with Wubeiling culture (97.2\%) and limited-to-moderate prior exposure to AI digital humans (97.2\%). This limited prior exposure reduces potential confounds, supporting the interpretability of both subjective experience measures and history knowledge assessments.

All participants provided written informed consent prior to the study, specifically agreeing to the temporary use of their facial and vocal data for stimuli generation. All personal data were strictly protected and deleted immediately after the experiment.

\begin{figure}[t]
  \centering
  \includegraphics[width=1\columnwidth]{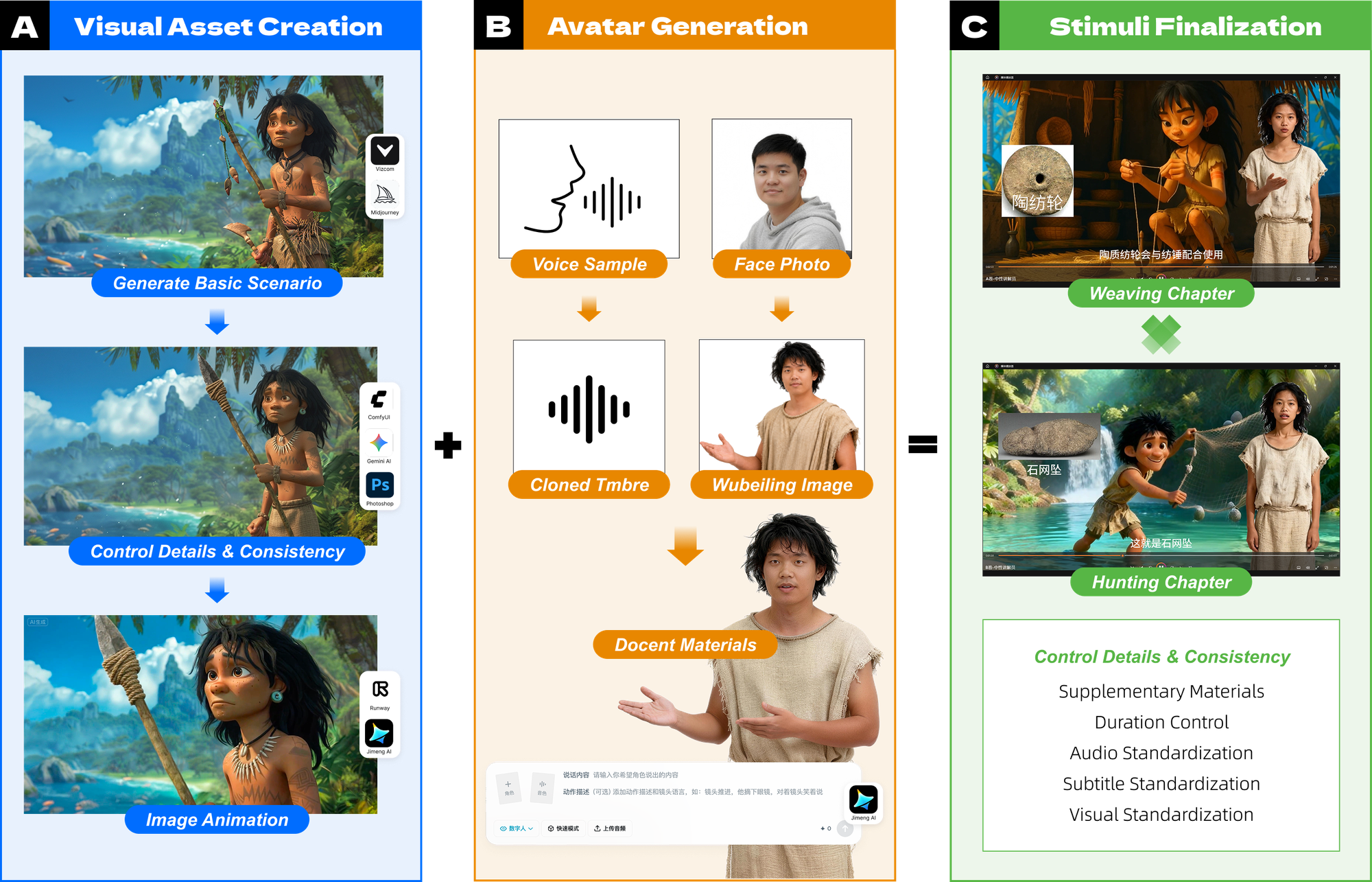}
  \caption{Overview of the stimuli development workflow.}
  \Description{Overview diagram of the three-step workflow used to create the digital self and non-self pedagogical agent video stimuli, including narrative design, asset creation, avatar generation, and compositing.}
  \label{fig:workflow}
\end{figure}

\subsection{Stimuli Development}

We developed the stimuli using first-hand research data and archaeological findings curated by Wubeiling archaeological specialists on our team. The learning content was structured around historically grounded narratives of Wubeiling culture and divided into two equivalent A/B sets with matched content volume. The stimuli were produced in three steps (Figure~\ref{fig:workflow}):


\textbf{(1) Visual narrative asset creation:} Because no definitive facial reconstruction of the Wubeiling people exists, we consulted archaeological experts and used an animated style to avoid over-specifying uncertain features. Generative AI was then used to create Wubeiling daily-life backgrounds set 3,000 years ago.

\textbf{(2) Multimodal narrator generation:} For the digital-self condition, participants provided a frontal photograph and a 10--15~s fixed-script voice recording for facial transfer and voice cloning. For the regular non-self agent condition, we used a standardized female character with expert-confirmed Lingnan region facial features. Both agents were generated through the same pipeline. 

\textbf{(3) Video composition and final processing:} We composited the docent footage with the narrative visuals, inserted supporting images for key historical references, and applied consistent subtitles to finalize the stimuli videos.

\subsection{Study Design}
We employed a counterbalanced within-subjects design, in which each participant experienced both agent conditions (digital-self and non-self) across two consecutive rounds. To control for potential confounds between learning content and condition, the two content sets were counterbalanced: one group viewed Content A with the digital-self and Content B with the non-self agent, while the other group received the reverse pairing. The presentation order was also counterbalanced, with half of the participants completing the digital self condition first and the other half starting with the non-self agent condition. The procedure was as follows:

\textbf{(1) Round 1:} Participants watched the first-round video, completed the corresponding learning experience questionnaires, and then took a content-specific knowledge quiz. 

\textbf{(2) Round 2:} Participants watched the second-round video and repeated the same sequence. 

\textbf{(3) Semi-structured interview:} Participants provided feedback on their learning experience and reasons for their preferences.

\subsection{Measures}

We assessed learning effectiveness using mixed quantitative and qualitative measures. For learning experience, we employed the Inclusion of Other in the Self (IOS) scale\cite{aron1992inclusion} for perceived psychological distance (IOS1: self--Wubeiling culture; IOS2: self--Wubeiling people in the background scenes), the Agent Persona Instrument (API)\cite{baylor2003api} for agent perception, the Narrative Transportation (NT) Scale\cite{green2000role} for immersion, the Intrinsic Motivation Inventory (IMI) relatedness subscale\cite{mcauley1989psychometric} for agent–user relatedness, and an Uncanny Valley scale (UVS)\cite{ho2017measuring} for negative affect. Learning outcomes were assessed using A/B-aligned quizzes, which were piloted with Wubeiling-unfamiliar individuals to ensure comparable difficulty, and the Remember/Know (R/K) paradigm\cite{gardiner1988functional} for memory state.

Qualitatively, we conducted brief, semi-structured interviews covering three themes: learning experience, agent relatedness, and educator expectations. All interviews were audio-recorded with participant consent and transcribed.

\section{Results}

\subsection{Quantitative Results}

\begingroup
\setlength{\intextsep}{6pt}      
\setlength{\abovecaptionskip}{2pt}
\setlength{\belowcaptionskip}{2pt}

\renewcommand{\arraystretch}{0.9} 

\begin{table}[H]
\caption{Quantitative analysis results ($N=36$).}
\label{tab:learning_measures_compact}
\centering
\footnotesize
\setlength{\tabcolsep}{4pt}
\renewcommand{\arraystretch}{1.08}

\begin{tabular*}{\columnwidth}{@{\extracolsep{\fill}} p{0.31\columnwidth} c c c c @{}}
\toprule
Measure & Self & Neutral & $p$ & ES \\
\midrule

IOS1
& 4.00 [3.00--5.00] 
& 2.00 [1.00--4.00] 
& $<.001$ 
& $r=.538$ \\

IOS2
& 4.00 [3.00--5.00] 
& 2.50 [2.00--3.75] 
& $.002$ 
& $r=.507$ \\

API
& 3.76 [3.49--4.00] 
& 3.34 [3.02--3.80] 
& $.027$ 
& $r=.370$ \\

NT
& 5.17 [4.54--5.67] 
& 4.67 [3.71--5.13] 
& $.012$ 
& $r=.417$ \\

IMI
& 5.01 (1.17) 
& 4.31 (1.18) 
& $.002$ 
& $d=.565$ \\

UVS
& 0.47 (0.71) 
& -0.73 (0.80) 
& $<.001$ 
& $d=.957$ \\

\midrule

History Retention
& 56.44 (14.47) 
& 63.50 (15.06) 
& $.018$ 
& $d=-.413$ \\

Remember
& 8.14 (4.74) 
& 8.86 (3.35) 
& $.343$ 
& $d=-.160$ \\

Know
& 6.58 (2.99) 
& 6.56 (2.60) 
& $.949$ 
& $d=.011$ \\

Guess
& 5.28 (3.85) 
& 4.58 (3.29) 
& $.279$ 
& $d=.183$ \\

\bottomrule
\end{tabular*}

\begin{flushleft}
\footnotesize
\textit{Note.} Wilcoxon rows report median [Q1--Q3] and effect size $r$; paired $t$-test rows report mean (SD) and Cohen's $d$. Two-tailed $p$ values are reported.
\end{flushleft}
\end{table}
\endgroup

For quantitative analysis, we used within-subject comparisons between the self and non-self conditions. Shapiro–Wilk tests were first used to assess the normality of difference scores, followed by paired-samples $t$-tests for approximately normal outcomes and Wilcoxon signed-rank tests for non-normal data. All tests were two-tailed ($\alpha = .05$), with results summarized in Table~\ref{tab:learning_measures_compact}.

Regarding the learning experience, the self condition increased several positive experiential measures compared with the non-self condition. Statistical analysis revealed significantly greater psychological closeness both to the Wubeiling culture (IOS1; $p < .001$) and to the Wubeiling people in the background scenes (IOS2; $p = .002$) under the self condition. The results also showed a more positive agent perception (API; $p = .027$), deeper narrative immersion (NT; $p = .012$), and stronger agent–user relatedness (IMI; $p = .002$). However, Uncanny Valley eeriness was also higher in the self condition, indicating increased negative affect (UVS; $p < .001$).

In terms of learning outcomes, the self condition did not yield higher performance. History retention scores were significantly lower under the self condition ($p = .018$), while memory-state measures (Remember, Know, or Guess) showed no reliable differences between the two conditions (all $p > .05$).

\subsection{Qualitative Findings}

For qualitative analysis, we adopted a collaborative inductive coding approach. Three researchers independently reviewed the transcripts, extracted salient excerpts (248 quotes), and generated initial codes. We then compared interpretations, clustered codes through affinity mapping, and resolved discrepancies through discussion before consolidating higher-order themes with representative quotes.

\textbf{\textit{Self-Similarity Evokes Familiarity Yet Can Also Cause Discomfort.}} Familiarity with the digital self enhanced users’ motivation to explore. As P31 noted: “Seeing my own face made me want to continue exploring.” Familiar sensory cues also lowered the processing threshold; P16 mentioned that a familiar voice “reduces the cognitive burden of identifying what is saying.” However, familiarity could trigger negative psychological defenses. P1 described a sense of social exposure: “[I felt] embarrassed when [my image] appeared on the screen.”

\textbf{\textit{Seeing a More Knowledgeable Self Can Be Motivating.}} Seeing the ancestral digital self behave beyond one’s current abilities could feel motivating. P30 noted, “Seeing my digital self has mastered this knowledge, I feel more confident about learning it myself.” This also enabled positive self-projection: P13 said it made him think he “might become this kind of presenter one day.” However, the discrepancy could also trigger discomfort. P5 said, “It is weird to see a familiar yet strange me doing things I have never done,” and P18 framed it as a loss of agency: “Digital self’s actions are entirely produced from data, and that’s outside my control.”

\textbf{\textit{Novelty Increases Attractiveness but Also Draws Attention Away from Learning Content.}} The digital self introduced a strong sense of novelty that increased attractiveness. As P23 said, “I find it really novel, so I just want to give it a try.” However, many participants reported allocating disproportionate attention to the digital self itself. P28 remarked, “It feels so magical—I can’t help staring at it all the time.” This attentional capture may come at the expense of the learning content; as P22 put it, “I remember the visuals, but I don’t remember what it actually talked about.”

\section{Discussion}

\subsection{Bridging History via the Digital Self}

Our results indicate that the digital self significantly enhanced multiple subjective dimensions of the learning experience, including narrative transportation, perceived relatedness, and self–other inclusion as well as agent perceptions—credibility, learning facilitation, engagement, and human-likeness. This aligns with the Self-Reference Effect, which suggests that self-related cues serve as salient cognitive anchors that facilitate deeper information processing \cite{rogers1977self,slater2009place}. The digital self functions not merely as a visual surrogate but as an identity mediator that bridges psychological distance \cite{MASSARA2013108} in temporally distant domains.

In this process, digital self may support historical empathy by enabling learners to establish a more personal and affective connection with historical narratives \cite{endacott2013updated}, such as projecting the AI video character as “my sister.” However, historical empathy and deeper reflection were not directly measured. Our findings therefore support perceived connection and experiential engagement rather than broader historical understanding.

Accordingly, when the design goal is to help learners enter a narrative and form emotional or imaginal connections, the digital self can serve as an effective design lever. This approach may also extend to other contexts with high perceived self-distance, such as cross-cultural learning or social-issue documentaries, though further research is needed.

\subsection{Decoupling of Learning Experience and Outcomes}

While the digital self significantly enhanced learners’ learning experience, it did not yield superior learning outcomes compared to the regular pedagogical agent. This suggests a potential decoupling between the quality of the learning experience and measurable learning outcome in short-term tasks.

Several factors may explain this discrepancy. One primary reason relates to the \textbf{novelty effect} \cite{poppenk2010revisiting}, where learners' initial attention is drawn more to the virtual avatar than to the instructional content. Consistent with previous findings \cite{herbert2024teaching}, self-avatar representations can capture learners’ attention and momentarily distract from the learning material, potentially offsetting gains in short-term performance. Another factor involves the \textbf{uncanny valley effect} \cite{mori2012uncanny}; higher perceived uncanniness increases adaptation time, further reducing immediate learning effectiveness \cite{10.1145/3652988.3673970}.

These findings suggest that identity salience should be calibrated to instructional goals rather than maximized. We propose three design strategies: \textbf{\textit{Adaptation Phase}}, gradually introducing the digital self to reduce novelty; \textbf{\textit{Selective Presence}}, emphasizing the agent at key moments while minimizing its presence during instruction; and \textbf{\textit{Stylized Abstraction}}, using non-photorealistic representations to reduce uncanniness while retaining identifiable features. These strategies aim to preserve experiential engagement without undermining immediate learning performance.

\section{Conclusion and Future Work}

We investigated an Ancestral Digital Self, a history-specific instantiation of the digital self, within history learning. Through a within-subjects study ($N=36$), we compared digital-self agents against non-self ones using AI-generated narrative videos. Results demonstrate that compared to the regular pedagogical agent, the digital Self enhanced experiential measures. However, it did not improve learning outcomes. This study highlights a critical trade-off between identity-driven engagement and knowledge acquisition, providing valuable design insights for balancing immersive self-representation with pedagogical effectiveness in future personalized learning environments. 

This approach could also be extended to other learning contexts. Future research should explore real-time interactive digital self systems to provide adaptive guidance and feedback, and conduct longitudinal studies to examine the sustained effects of digital selves on motivation, immersion, and knowledge retention.

\bibliographystyle{ACM-Reference-Format}
\bibliography{references}

\end{document}